\documentclass[10pt,preprint] {aastex}

\usepackage{graphicx,epsfig,natbib,color}
\usepackage{lscape}
\usepackage{graphicx}
\usepackage{epsfig}
\usepackage{natbib}

\begin{document}
\title{Observation of Magnetic reconnection at a 3D null point associated with a solar eruption}
\author{J. Q. Sun\altaffilmark{1,2,3}, J. Zhang\altaffilmark{3}, K. Yang\altaffilmark{1,2}, X. Cheng\altaffilmark{1,2}, M. D. Ding\altaffilmark{1,2}}
\affil{$^1$ School of Astronomy and Space Science, Nanjing University, Nanjing 210023, China}
\affil{$^2$ Key Laboratory for Modern Astronomy and Astrophysics (Nanjing University), Ministry of Education, Nanjing 210023, China}
\affil{$^3$ Department of Physics, Astronomy and Computational Sciences, George Mason University, 4400 University Drive, MSN 6A2, Fairfax, Virginia 22030, USA}
\email{dmd@nju.edu.cn; jzhang7@gmu.edu}

\begin{abstract}
Magnetic null has long been recognized as a special structure serving as a preferential site for magnetic reconnection (MR). However, the direct observational study of MR at null-points is largely lacking. Here, we show the observations of MR around a magnetic null associated with an eruption that resulted in an M1.7 flare and a coronal mass ejection. The GOES X-ray profile of the flare exhibited two peaks at $\sim$02:23 UT and $\sim$02:40 UT on 2012 November 8, respectively. Based on the imaging observations, we find that the first and also primary X-ray peak was originated from MR in the current sheet underneath the erupting magnetic flux rope (MFR). On the other hand, the second and also weaker X-ray peak was caused by MR around a null-point located above the pre-eruption MFR. The interaction of the null-point and the erupting MFR can be described as a two-step process. During the first step, the erupting and fast expanding MFR passed through the null-point, resulting in a significant displacement of the magnetic field surrounding the null. During the second step, the displaced magnetic field started to move back, resulting in a converging inflow and subsequently the MR around the null. The null-point reconnection is a different process from the current sheet reconnection in this flare; the latter is the cause of the main peak of the flare, while the former is the cause of the secondary peak of the flare and the conspicuous high-lying cusp structure. 
\end{abstract}
Online-only material: animations, color figures

\section{INTRODUCTION}
Magnetic reconnection (MR) has long been recognized as the principal physical mechanism responsible for solar flares. Where and how it takes place remains so essential in understanding the triggering and evolution of flares and associated coronal mass ejections (CMEs). The dual phenomena of flares and CMEs can be collectively called a solar eruption, because of close time coincidence between the impulsive phase of flares and strong acceleration phase of CMEs, both of which are likely driven by MR \citep{Zhang2001a}. Over the past several decades, the flare-producing MR has been intensively studied and succinctly described with a standard `CSHKP' flare model \citep{Svestka1992a} (CSHKP refers to the studies of \citealt{Carmichael1964a, Sturrock1966a, Hirayama1974a, Kopp1976a}). This model, consisting of a magnetic flux rope (MFR) and an underneath current sheet (CS) \citep{Lin2000a}, has been largely validated by many observational studies that showed consistent features, such as plasma inflows \citep{Yokoyama2001a, Sun2013a, Sun2015a, Zhu2016a} , outflows or down flows \citep{McKenzie1999a, Liu2013a}, loop-top hard X-ray sources\citep{Masuda1994a}, coronal X-ray sources \citep{Sui2003a}, the termination shock \citep{Chen2015a}, and so on. The CSHKP model has been successful in explaining a large array of observational features associated with eruptive flares. Nevertheless, it is largely a two-dimensional (2D) model and has a simple bipolar magnetic configuration. It has an intrinsic shortage in describing certain realistic solar flare events, which may explicitly involve the three-dimensional (3D) structure in the modeling in order to understand certain observational features.\par

One of the important subjects in 3D MR theories concerns the peculiar structure of null-points. Previous theoretical studies suggest that 3D null-point may be a preferential site for the generation of MR \citep{Priest1996a, Priest2009a}. Based on the theoretical models, a series of numerical experiments are conducted through applying perturbations, like rotational and shearing motions, to the spine and fan structures extended from the null. As a result, strong current is found to develop around the null-point \citep{Pontin2007a, Galsgaard2011a, Wyper2012a}. \cite{Pontin2013a} demonstrated that the connectivity of the magnetic field line would change when it transfers across the CS forming around the null-point and treat that as a feature of spine-fan reconnection. In addition, the role of 3D reconnecting null-point in accelerating particles is also confirmed by several simulations \citep{Browning2010a, Stanier2012a, Baumann2013a, Baumann2013b}. However, in observations, the reconnection around null-point has only been reported in a handful of studies in the Sun \citep{Filippov1999a, Fletcher2001a, Sun2012a, Wang2012a, Sun2014a} and in the Earth's magnetosphere \citep{Xiao2006a}. The detailed evolution of the magnetic field and plasma surrounding the null-point during the reconnection, i.e. the observational features of the null-point reconnection, has been seldom obtained. Such lack of observational studies restrict the improvement of theoretical models. Consequently, how the null-point reconnection is triggered and conversely how it affects the evolution of flares are still open questions that need more observations to answer \citep{Torok2011a}.\par

Here, we show the direct observations that reveal the processes of how MR is triggered and how it proceeds at a null-point. The observations are based on an array of instruments, including the Solar Dynamics Observatory (SDO; \citealt{Pesnell2012a}), Reuven Ramaty High Energy Solar Spectroscopic Imager (RHESSI; \citealt{Lin2002a}), Nobeyama Radioheliograph (NoRH; \citealt{Nakajima1994a}), and Geostationary Operational Environmental Satellites (GOES). The overall observations are presented in $\S$2.1. $\S$2.2 focuses on the process of MR, followed by the analysis of the 3D magnetic topology in $\S$2.3. We conclude the results and discuss their significance in $\S$3.\par

\section{Results}
\subsection{Observations of the flare} 
On 2012 November 8, a GOES M1.7-class flare occurred near the northeastern solar limb. From the GOES 1-8 {\AA} soft X-ray (SXR) flux profile, the flare started at $\sim$02:10 UT and peaked at $\sim$02:23 UT. The gradual declining of the flux was interrupted by a second enhancement at $\sim$02:38 UT and reached a second peak at $\sim$02:40 UT (Figure 2(c)). This double-peak property of the flare emission is more obvious in the profiles of Atmospheric Imaging Assembly (AIA; \citealt{Lemen2012a}) 94 {\AA} and 131{\AA} fluxes (Figure 2(c)). Prior to the SXR increase, a filament already began to rise up slowly, producing some sporadic brightenings along the filament structure (Figure 1(a)-(b)). As the slow-rising motion of the filament continued, a 6-12 keV X-ray source appeared slightly underneath it at 02:08 UT and was quickly evolved to contain higher energy emission of 12-25 keV at 02:10 UT. About one minute later, probably due to plasma heating, the upper part of the filament disappeared in 304 {\AA} and 171{\AA} images; instead, an `M' shaped hot structure appeared there in 131 {\AA} passband (Figure 1(c)). This hot structure is likely to be the core part of the erupting MFR. Under the MFR, a post-flare loop arcade appears in 131 {\AA} passband (Figure 1(c)-(d)). At the same time, the 25-50 keV non-thermal HXR sources resulting from the accelerated electrons are found at two footpoints and above the top of the loop. These observational features are consistent with the classical flare models that MR takes place in a vertical CS located between the CME and the post-flare loops.\par

Following the eruption, the flare soon achieved the SXR peak at $\sim$02:23 UT and evolved into the gradual phase. However, the gradual decay of the SXR emission just lasted for a short time; 15 minutes later, at $\sim$02:38 UT, the SXR began to increase again and reached a second peak at $\sim$02:40 UT. Correspondingly, there also existed a two-stage energy release in HXR, appearing as two independent bumps in Fermi HXR spectrogram (Figure 2(a)). The second bump appeared after 02:35 UT and was less energetic (mainly under 20 keV) than the first one in the impulsive phase (up to more than 100 keV). In AIA 193 {\AA} images, a new southeast-directed cusp appeared at a higher altitude after 02:40 UT (Figure 1(f)). We chose two slices, i.e. Slice 1 and Slice 2 as plotted in Figure 2(d), along and perpendicular to the cusp, respectively, to extract the time-distance evolution maps of the cusp. Along the cusp, we find two above-the-loop-top hot regions ($\geq$10 MK) (Figure 2(b)) that appeared successively in time, coinciding temporally with the SXR double peaks. Hot region 1 appeared in the impulsive phase of the flare at a height of $\sim$10 Mm. Prior to its disappearance at $\sim$02:36 UT, it continuously moved upward for $\sim$10 Mm. Hot region 2 emerged at a higher altitude of $\sim$22 Mm. However, the location of hot region 2 almost remained stationary during its $\sim$5 minutes of apparent life. All the observations from EUV to HXR show the evidence of a second energy release, which originated from a different location and had different properties from the one in the first and main peak of the flare.\par

\subsection{The 3D null-point magnetic reconnection}
In the impulsive phase, the main features of the flare, including the outward eruption bubble, the underneath flare loop arcade, the coronal and footpoint HXR sources, and the rising above-the-loop-top hot region, can be explained by the CSHKP model. Due to the ascent of the presumed MFR, the reconnection site in the vertical CS is also expected to move upward, causing the rise of the flare loop arcade and the hot region 1 (Figure 2(b)).\par 
However, the CSHKP model could not explain the second peak, which appeared in the gradual phase. In $\S$2.1, we have already shown some evidence of MR responsible for the second peak, such as a higher southeast-directed cusp, an above-the-loop-top hot region as well as the emission enhancement from EUV to HXR. In the following, we focus on some dynamic properties of the second-peak reconnection, i.e. where and how magnetic fields reconnect. It is found that the evolution of the magnetic field participating in such a reconnection can be divided into two steps: (1) the separation distortion, and (2) the recovery inflow leading to the MR. The outward-moving and expanding MFR distorted the surrounding coronal fields, in particular, caused a large separation displacement of the overlying magnetic fields along the direction of slice 2 from 02:10 UT to 02:20 UT (Figure 3(d)). The departure of the MFR from the flare region left an area of low magnetic pressure behind it. Due to the required pressure balance with the surroundings, the corona loops pushed aside during the eruption, denoted as `Coronal loops S' (CLS) and `Coronal loops N' (CLN) in Figure 1(g), began to recover and approach each other; the approaching or inflow velocity is in a range of 1-10 km s$^{-1}$ at $\sim$03:03 UT, calculated from the Fourier local correlation tracking (FLCT) method (Figure 3(a)). The inflow formed an X-shaped structure seen in 171 {\AA} images. According to the frozen-in property in a highly conductive plasma, these observed loops may well trace out the magnetic field lines in the corona, thereby indicating an X-shaped magnetic topology there. Once the CLS and CLN contacted with each other at the X-point (the center of the X-shaped structure, marked by an orange cross in Figure 3(b)-(c)), they suddenly disappeared (Figure 3(d)). Immediately afterward, a cusp structure (Figure 1(h)) was formed and lightened up in the 131 {\AA} passband under the X-point. With the AIA multi-wavelength observations, we calculate the differential emission measure (DEM) using the method from \cite{Hannah2012a}. We define the total emission measure (EM) and DEM-weighted temperature (T) as:
\begin{equation} 
 EM  =  \int _{T_{\mathrm{min}}} ^{T_{\mathrm{max}}} DEM(T) \mathrm{d}T,
 \end{equation}
 and
\begin{equation}T_{\mathrm{mean}} = \frac{ \int_{T_{\mathrm{min}}}^{T{\mathrm{max}}} DEM(T) T \mathrm{d}T} {\int_{T {\mathrm{min}}}^{T{\mathrm{max}}} DEM(T) \mathrm{d}T}.   \end{equation}
The EM and temperature distribution maps of the flare region at 03:03 UT are plotted in Figure 3(b)-(c). It is seen that the hottest region in the cusp structure was located slightly lower than the cusp tip point, the presumed location of the MR, but higher than the above-the-loop-top HXR source. The center of the above-the-loop-top HXR source was about 5$\arcsec$ higher than the highest EM region. From the cusp tip to the hottest region, both the temperature and EM increased significantly. The hot cusp-like loops in the region, which were newly generated at the tip of the cusp through MR, had apparently a downward contracting motion, which is also know as flare loop shrinkage. Meanwhile, many downward small scale outflows appeared at the cusp tip in the 94 {\AA} and 131 {\AA} passbands (Figure 3(e)). Most of these outflows moved downward with initial velocities of $\sim$50--150 km s$^{-1}$ and then stopped as they approached the top of the loop arcade. Notably, the outflows after 02:40 UT almost appeared to start at the similar height in the corona (Figure 3(e)). The dynamical evolution characteristics of the cool and hot plasma support the scenario of the MR concisely situated at the X-point.\par

Underneath the outflow region, there successively appeared a coronal 12-25 keV X-ray source and a 34 GHz radio source, which are thought to be radiated from energetic electrons via bremsstrahlung and cyclotron emission mechanism, respectively. The energetic electrons streamed down along field lines, and bombarded the chromosphere, forming the brightenings seen in 1600 {\AA} and 1700 {\AA} passbands (Figure 3(a)). These observations further confirm the existence of MR above the cusp. It is worthy noting that this MR process is different from that during the main flare peak. Topologically, CLS and CLN were apparently impossible to participate in the MR underneath the MFR responsible for the main peak, since they were located above and in the periphery of the original MFR. Besides, the differences in the occurrence height and orientation between the post-flare arcade and the cusp also mean that the MR at the main peak and that at the second peak must take place at different locations. At the end of the flare, there appeared two groups of post-flare loop arcades at different heights (Figure 1(i)), which were the results of the two episodes of MR, respectively.\par

\subsection{Three-dimensional topology of the magnetic reconnection at the magnetic null-point}
We have shown above the observational evidence of two-episode MR in a flare process. To further understand how MR takes place, we extrapolate the 3D coronal magnetic field based on the photospheric vector field observed by Helioseismic and Magnetic Imager (HMI; \citealt{Scherrer2012a, Schou2012a}) using a NLFFF model. The data are pre-processed into the Space-weather HMI Active Region Patches (SHARPs) heliographic Cylindrical Equal-Area (CEA) form \citep{Bobra2014a}. Since the active region is rather quiet (no C-class flares and above) during the following two days after the flare, we use the vector magnetic field at 12:00:00 UT on 2012 November 10 (HARP 2186) as the bottom boundary for the extrapolation. The vector magnetic field of the active region shows a positive parasitic polarity (P1) encircled by three negative polarities located in the north, southeast, and west sides, respectively (N1, N2, and N3, in Figure 4(a)). Such a distribution is analogous to the lower surface of the magnetic `dome' topology in \cite{Pontin2013a}. Through the NLFFF extrapolation, we also derive a similar dome-like structure, that is composed of the fan and spine structures associated with a null-point (Figure 4 (c)). Interestingly, the height of the null-point is $\sim$60 Mm over the photosphere, almost the same as the height of the reconnection site of the second peak. Such a spatial coincidence indicates that the second-peak MR likely occurs around the magnetic null. We compare the magnetic field lines with the 171 {\AA} coronal loops around the cusp (Figure 4(c)-(d)) and find that they are mostly similar in topological configuration. Specifically, the CLS corresponds to one group of closed field lines connecting P1 and N2 (green lines in Figure 4(c)). The CLN has an identical location and orientation to the open field lines originating from N1 and passing nearby the null-point (yellow lines in Figure 4(c)). Both groups of field lines have sharp kink when they approach the magnetic null, outlining an X-shaped structure in projection, consistent with the 171 {\AA} X-shaped structure (Figure 1(g)). When they cross the magnetic null, their connectivities are supposed to change dramatically; in other words, magnetic field lines can `reconnect' there. As a result, a new group of magnetic field lines -- newly reconnected field lines are produced, which are, in EUV observations, the field lines shaping the hot cusp structure (red lines in Figure 4(c)). Since the reconnection takes place between the field lines in the spine and fan (Figure 4(c)), it probably corresponds to the spine-fan reconnection as \cite{Priest1996a} and \cite{Pontin2013a} defined.\par

\section{Discussion and conclusion}
In this work, we analyze the MR around a null-point in an M1.7 flare. The main results can be summarized as follows:
 
\begin{enumerate}
\item{There existed two separate peaks in SXR and HXR time profiles resulted from MR in two different locations: one from the vertical CS created by the erupting MFR, and the other probably from the null-point roughly stationed above the pre-eruption MFR.}
\item{The magnetic field associated with the null-point went through two distinct steps: (1) separation distortion due to the throughput motion of the eruption/expanding MFR, and (2) inflow or recovery motion of the displaced field lines following the eruption.} 
\end{enumerate}\par

Based on the above results, the two-step MR during the flare is schematically shown in Figure 5. During the flare impulsive phase, the MR (red star) takes place in the CS underneath the erupting MFR (red field lines) and produces the flare arcade 1(blue loops). As the MFR erupts through the MN, the magnetic field surrounding the null (green and yellow field lines) are pushed aside. When the MFR moves away, these distorted field lines begin to move back and produce the MR around the MN, produce the cusp (orange region) and flare arcade 2 (red loop).\par

In classical flare models that involve the current sheet, the reconnection site always keeps rising with time, due to the upward erupting MFR leading to the further stretching of the overlying magnetic field to a higher altitude. However, during the null-point reconnection, both the evolution of the above-the-loop-top hot region and the outflows indicated that the reconnection site (i.e. the magnetic null) almost keeps at the same altitude. However, the null-point reconnection also shows many similarities to the traditional flare reconnection, including the plasma inflow toward and outflow from the reconnection site, the flare cusp, the above-the-loop-top hot region, the HXR coronal source, and the post-flare loop arcade etc. Moreover, from the emission feature of the cusp, the most intensive emission region also lies somewhere outside the MR site \citep{Krucker2010a, Liu2013a, Sun2014b}. 

Since magnetic nulls are rather common in the corona, they have the potential to trigger the reconnection there when encountered by an erupting MFR \citep{Torok2011a}, like what we observed in this event. This kind of MR may not be uncommon in solar flares, but is largely ignored in previous studies, mainly limited by observations. \par

Due to uncertainties in the magnetic field measurements, which are used as the bottom boundary, and the NLFFF modeling, we cannot completely rule out other MR regimes at flare second peak. However, based on the above observations, we believe the null-point reconnection is a reasonable explanation. More features need to be explored theoretically and observationally to uncover the mechanism of the null-point reconnection in the future.

 \acknowledgements The authors thank the referee for constructive comments that helped improve the paper. This work was supported by NSFC under grants 11303016, 11373023, and NKBRSF under grant 2014CB744203. J. Zhang was supported by NSF AGS-1249270 and AGS-1460188.


\begin{figure*} 
      \vspace{-0.0\textwidth}    
      \centerline{\hspace*{0.00\textwidth}
      \includegraphics[width=1.0\textwidth,clip=]{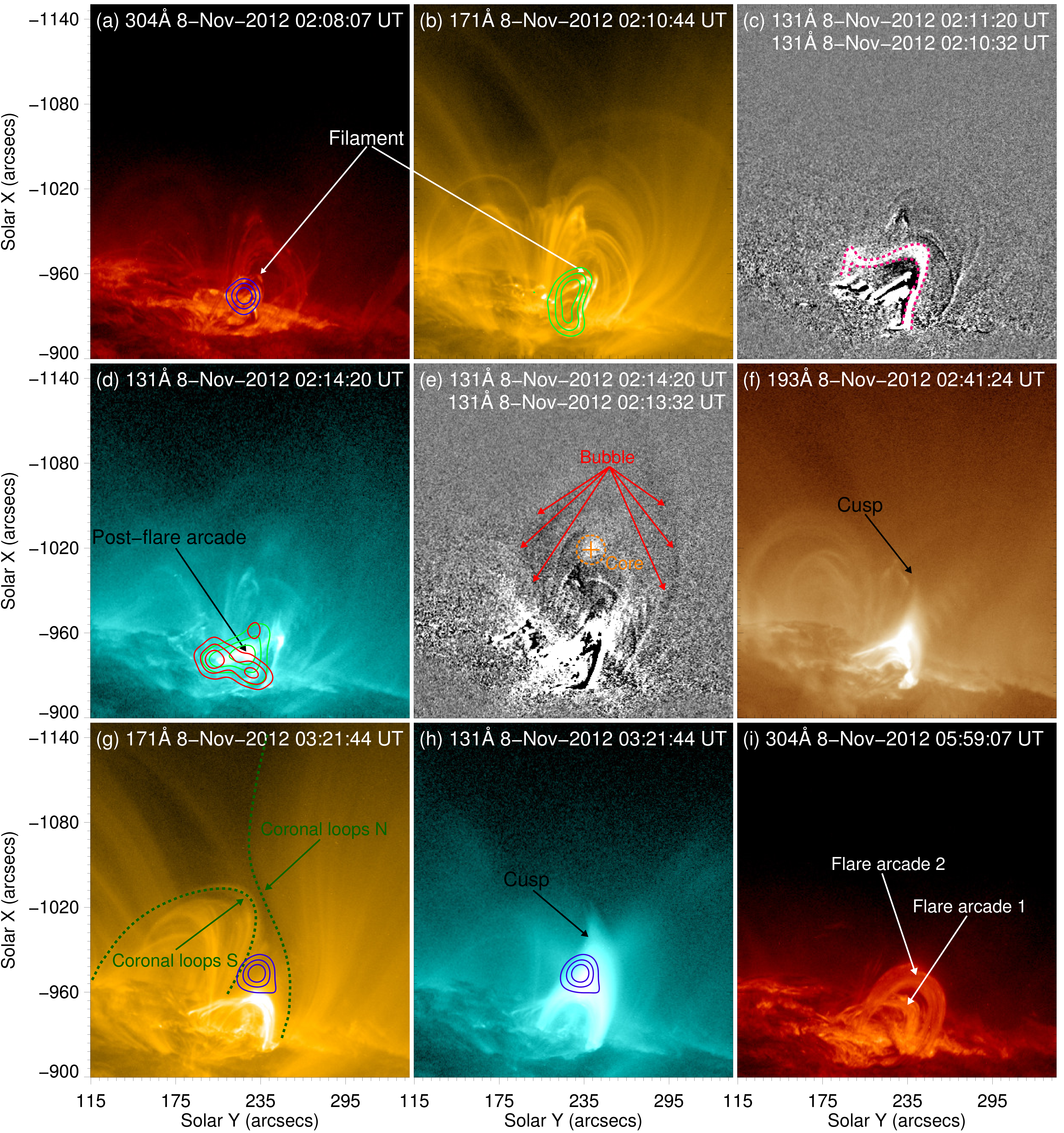}
      }
\caption{ \textbf{Overview of the flare evolution.} (a)-(b) AIA 304 {\AA} and 171 {\AA} images of the flare region at 02:08:07 UT and 02:10:44 UT, respectively. (c) AIA 131 {\AA} difference map between 02:11:20 UT and 02:10:32 UT. The pink dashed lines sketch a possible rising MFR. (d) AIA 131 {\AA} image of the flare region at 02:14:20 UT. (e) AIA 131 {\AA} difference map between 02:14:20 UT and 02:13:32 UT, showing an intense core surrounded by a bubble. (f) AIA 193 {\AA} image of the flare region at 02:41:24 UT, revealing a high-lying cusp. (g) AIA 171 {\AA} image of the flare region at 03:21:44 UT, the two cyan dashed lines sketch two groups of coronal loops that probably participating in the second-peak reconnection. (h) AIA 131 {\AA} image at 03:21:44 UT showing the cusp. (i) AIA 304 {\AA} image at 05:59:07 UT. The over-plotted contours are RHESSI 6-12 keV (blue), 12-25 keV (green), and 25-50 keV (red) X-ray sources at levels of 40\%, 60\%, and 80\%.} \label{f1}
(Animations are available in the online journal.)
\end{figure*}

\begin{figure*} 
     \vspace{-0.0\textwidth}    
     \centerline{\hspace*{0.00\textwidth}
               \includegraphics[width=1.\textwidth,clip=]{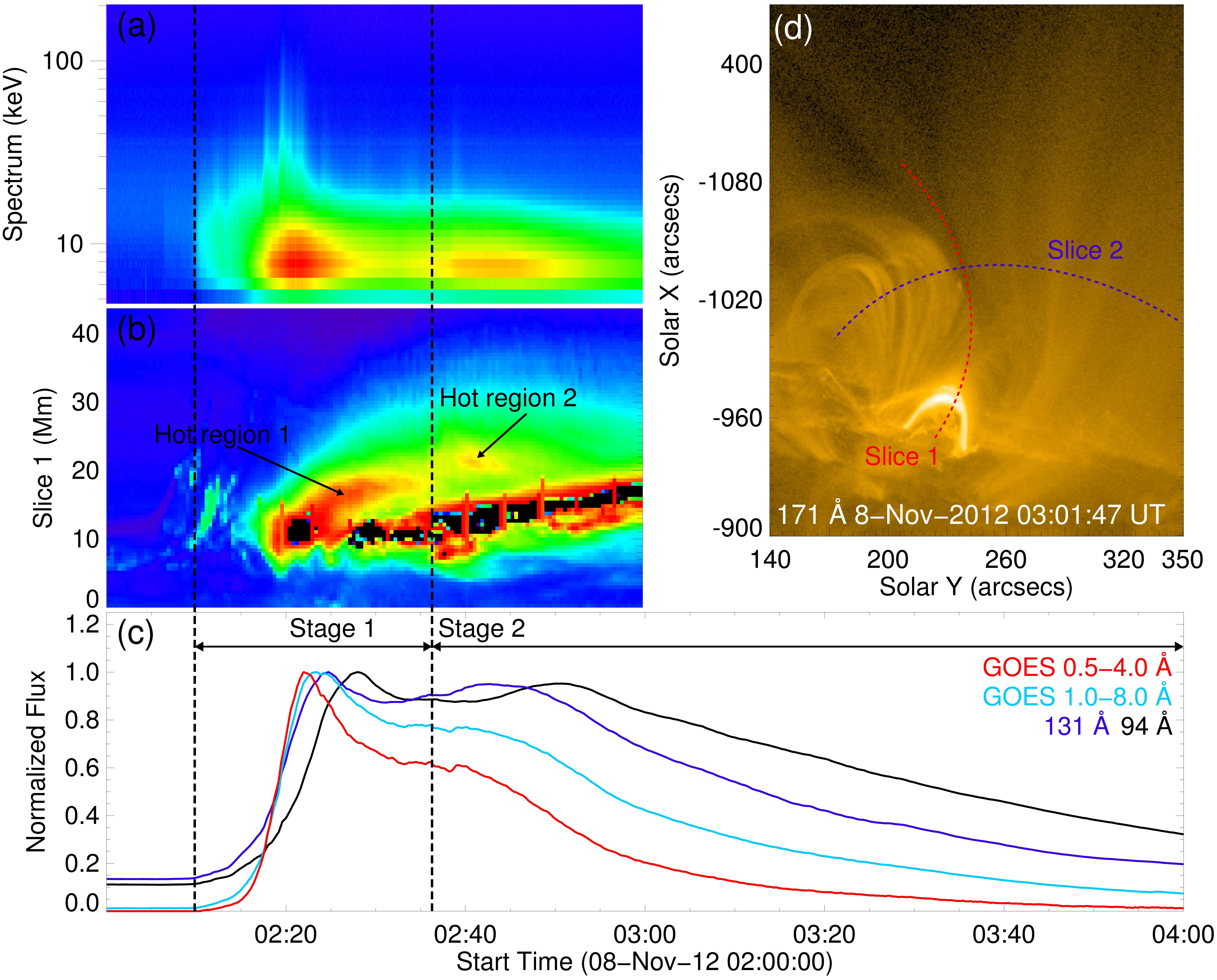}
               }
\caption{\textbf{Two episodes of energy release in the flare.} (a) Fermi 4.2-200 keV spectrogram. (b) Time-slice evolution of 193 {\AA} emission along the flare cusp. (c) Time profiles of the GOES SXR, AIA 94 {\AA}, and 131 {\AA} flux of the flare region normalized to the maximum values. (d) AIA 171 {\AA} image of the flare region at 03:01:47 UT. The blue and red dashed lines denote the slice 1 and 2 used for Figures 3(b), 4(d)-(e). The black dashed lines separate the flare evolution into two stages.} \label{f2}
\end{figure*}

\begin{figure*} 
     \vspace{-0.0\textwidth}    
     \centerline{\hspace*{0.00\textwidth}
               \includegraphics[width=1.15\textwidth,clip=]{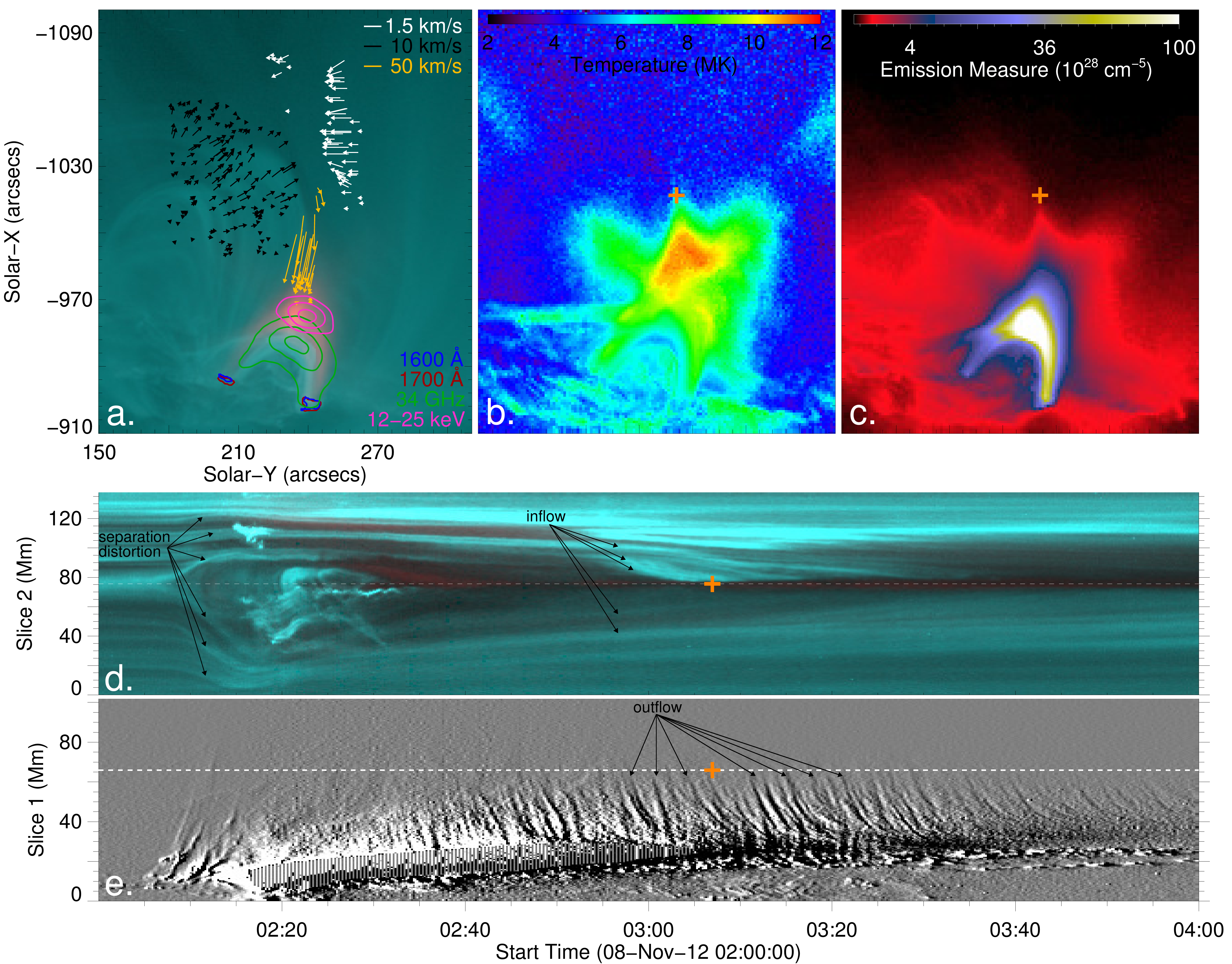}
               }
\caption{\textbf{Observational evidence of MR around a 3D null-point.} (a) A composite image of the AIA 131 {\AA} (red) and 171 {\AA} (cyan) passbands at $\sim$03:03 UT. The arrows scaled by the white, black, and orange line segments are inflow and outflow plasma velocities around 03:03 UT and 03:17 UT, respectively. The over-plotted contours in pink, green, blue, and red indicate the emission of 12-25 keV (50\%, 70\%, and 90\%), 34 GHz (30\%, 60\%, and 90\%), 1600 {\AA} (50\%), and 1700 {\AA} (50\%) around 03:03 UT, respectively. (b) DEM-weighted temperature map (c) EM map of the flare region. (d) 131 {\AA} (red) and 171 {\AA} (cyan) time evolution along Slice 2. (e) 131 {\AA} time-difference evolution along Slice 1. The orange crosses in panel (b)-(e) denote the same position, which is the suspected reconnection site. } \label{f3}
\end{figure*}

\begin{figure*} 
     \vspace{-0.0\textwidth}    
     \centerline{\hspace*{0.00\textwidth}
               \includegraphics[width=1.0\textwidth,clip=]{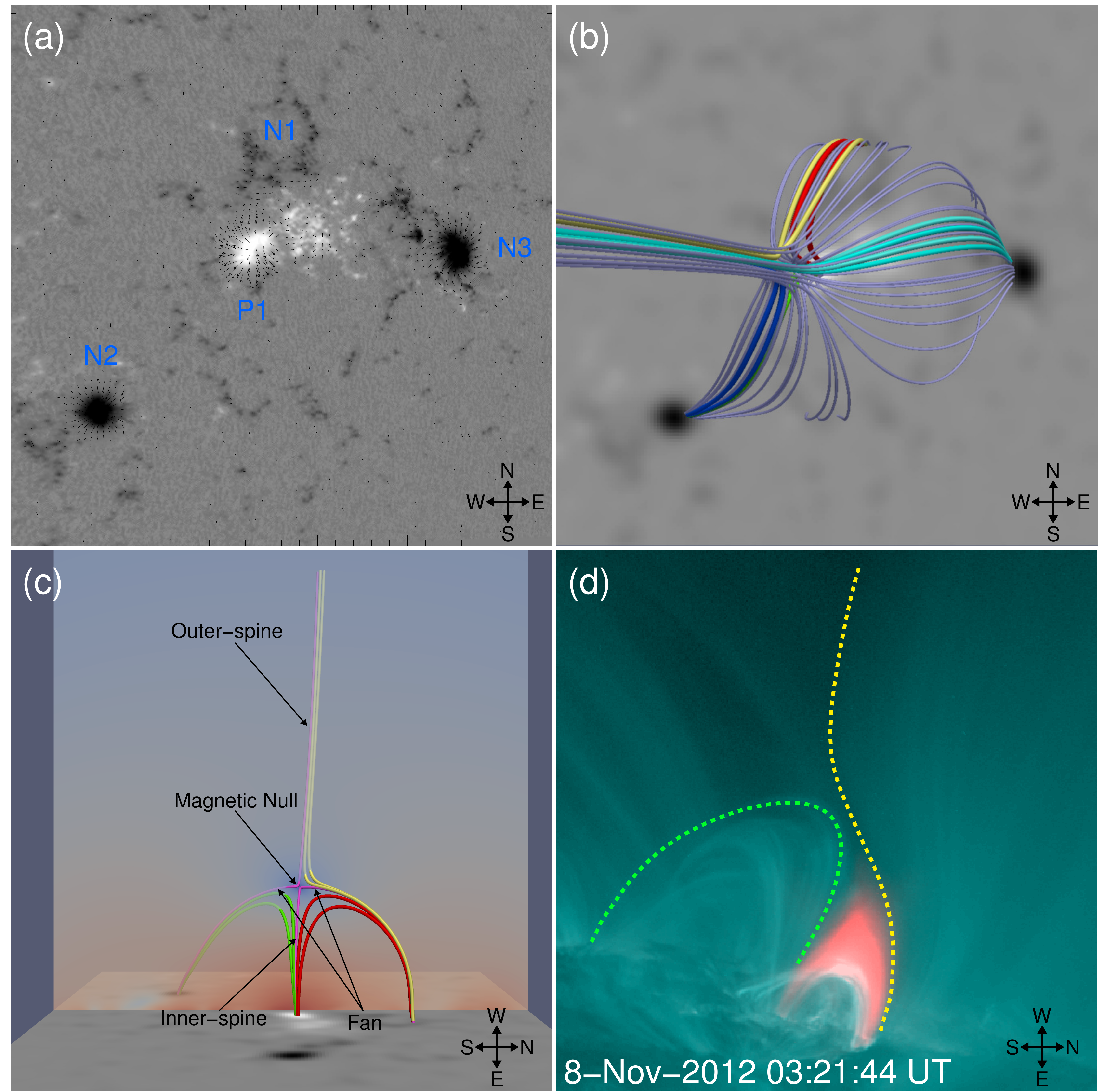}
               }
\caption{\textbf{The magnetic field topology of the active region.} (a) Photosphere magnetic field of the active region. `N1', `N2', `N3', and `P1' denote three negative polarities and one positive polarity, respectively. (b) A top view of the magnetic field lines in the active region. (c) Magnetic fields participating in the second peak null-point reconnection. The vertical cut showing the magnetic intensity map in which a minimum region corresponds to a magnetic null. The spine and fan lines are denoted and plotted in pink. (d) The AIA 131 {\AA} and 171 {\AA} composite image. The green and yellow dashed lines sketch the coronal loops participate in the null-point reconnection, which are also used to show a similarity to the green and yellow field lines in panel (c).} \label{f4}
(Animation is available in the online journal.)
\end{figure*}

\begin{figure*} 
     \vspace{-0.0\textwidth}    
     \centerline{\hspace*{0.00\textwidth}
               \includegraphics[width=1.\textwidth,clip=]{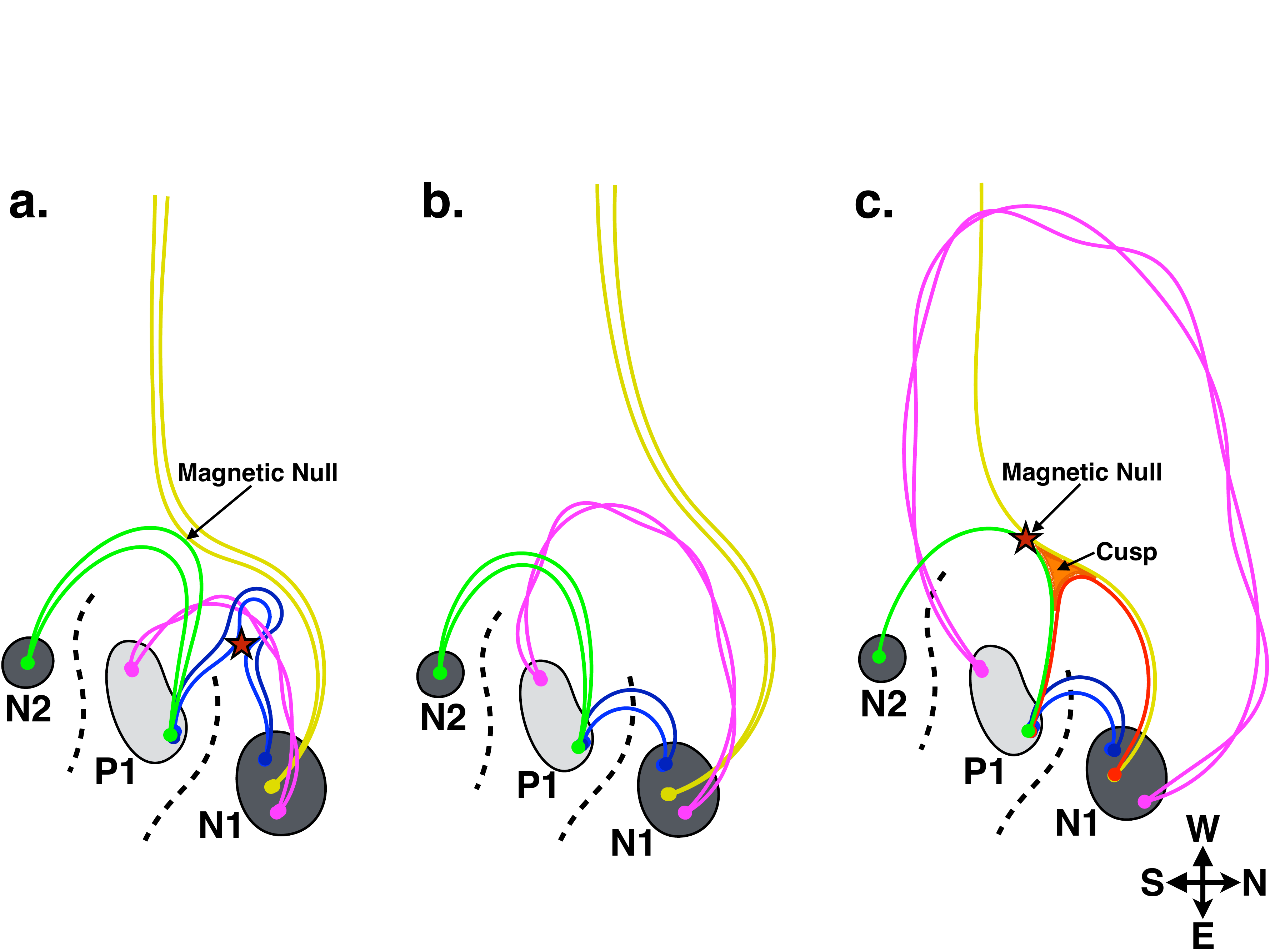}
               }
\caption{\textbf{Schematic drawing of the two-step MR involved in the dual-peak flares.} MR in the CS under the MFR is triggered when the MFR erupts upward (a). Meanwhile, the MFR distorts and separates the magnetic fields around the null-point (b), which, however begin to recover toward the null when the MFR leaves away and further causes the reconnection around the null (c).} \label{f5}
\end{figure*}

\end{document}